\documentclass[conference]{IEEEtran}
\IEEEoverridecommandlockouts

\usepackage{cite}
\usepackage{amsmath,amssymb,amsfonts}
\usepackage{acronym}
\usepackage{subfigure} 
\usepackage{multirow}
\usepackage{booktabs}
\usepackage{algorithmic}
\usepackage{graphicx}
\usepackage{textcomp}
\usepackage{xcolor}

\def\BibTeX{{\rm B\kern-.05em{\sc i\kern-.025em b}\kern-.08em
    T\kern-.1667em\lower.7ex\hbox{E}\kern-.125emX}}

\input{AcronymList}
\begin{document}

\title{Enhancing Channel Shortening Based Physical Layer Security Using Coordinated Multipoint\\ 
\thanks{This work is supported by the Scientific and Technological Research Council of Turkey (TUBITAK) under Grant No. 120C142.\\
This work has been submitted to the IEEE for possible publication. Copyright may be transferred without notice, after which this version may
no longer be accessible.}}

\author{
\IEEEauthorblockN{Muhammad~Sohaib~J.~Solaija\IEEEauthorrefmark{1}, Hanadi Salman\IEEEauthorrefmark{1}, and H\"{u}seyin Arslan\IEEEauthorrefmark{1}\IEEEauthorrefmark{2}}\\
\IEEEauthorblockA{\IEEEauthorrefmark{1}Department of Electrical and Electronics Engineering, Istanbul Medipol University, Istanbul, 34810 Turkey\\
}
\IEEEauthorblockA{\IEEEauthorrefmark{2}Department of Electrical Engineering, University of South Florida, Tampa, FL 33620 USA\\
Email: \{muhammad.solaija, hanadi.suleiman\}@std.medipol.edu.tr, huseyinarslan@medipol.edu.tr}
}

\maketitle

\begin{abstract}
Wireless networks have become imperative in all areas of human life. As such, one of the most critical concerns in next-generation networks is ensuring the security and privacy of user data/communication. Cryptography has been conventionally used to tackle this, but it may not be scalable (in terms of key exchange and management) with the increasingly heterogeneous network deployments. Physical layer security (PLS) provides a promising alternative, but struggles when an attacker boasts a better wireless channel as compared to the legitimate user. This work leverages the coordinated multipoint concept and its distributed transmission points, in conjunction with channel shortening, to address this problem. Results show significant degradation of the bit-error-rate experienced at the eavesdropper as compared to state-of-the-art channel shortening-based PLS methods. 

\end{abstract}


\section{Introduction}
\label{Sec:Intro}
Wireless communication has pervaded all aspects of human existence. The recent pandemic has further cemented its importance, with different facets of our daily lives including (but not limited to) education, retail, banking, healthcare all depending heavily on reliable wireless communication for their continuity despite the challenging restrictions worldwide \cite{ahmadi2020pandemic}. While this ubiquitous availability of wireless signals is desirable from a communication (and sensing) perspective, it is becoming increasingly challenging to ensure the privacy of confidential data and information when it is being transmitted openly into the environment \cite{furqan2021wireless}. This broadcast nature of wireless communication renders it susceptible to threats such as eavesdropping, jamming, and spoofing. In eavesdropping, the attacker tries to intercept and interpret the ongoing communication between legitimate nodes. In jamming, the attacker's target is to disrupt the communication, while in spoofing, the attacker impersonates a legitimate node for malicious purposes \cite{zou2016survey}. The latter two, though more serious in impact, are often preceded by the former. Presently, there are two well-established approaches for securing wireless networks, i.e., \textit{cryptography} and \textit{\ac{PLS}}. Cryptography secures the message or content of communication using keys for both encryption and decryption of the messages. However, this requires the exchange of keys (in symmetric encryption), their management, and computational capabilities at the terminals. \ac{PLS} provides a complementary solution for securing communication by protecting the wireless link (and the physical signal traversing it) between legitimate nodes using the properties of the channel and transceivers \cite{hamamreh2018classifications}.

In the context of this work, we focus on the development of a \ac{CoMP}-assisted \ac{PLS} solution, that leverages channel shortening to protect communication against eavesdropping attacks. While the \ac{CoMP} concept has been around for over a decade, only a handful of works have considered its application for securing wireless communication \cite{yusuf2015secure, hafez2017secure, wang2016signal,xu2016enhancing,yao20193d}. \ac{CoMP} has been leveraged in \cite{yusuf2015secure} to address the limitation of directional modulation as it fails to secure the communication if the eavesdropper is in the same direction as the legitimate receiver. The coordinating \acp{TP} transmit copies of the same signal using directional modulation, such that these are correctly received only at the legitimate receiver's location, providing location-based security. Building on top of that, \ac{CoMP} is proposed to be used for sparse radio environments such that data is only decodable at the intersection of information beams while a distorted constellation is observed at other locations \cite{hafez2017secure}. \ac{CoMP}-based transmissions from distributed antenna elements in an underwater scenario are proposed in \cite{wang2016signal}, where the power and schedule of transmissions are manipulated such that the messages are non-overlapping at the legitimate receiver while being overlapped (and interfering) at the eavesdropper. A dynamic \ac{CoMP} scheme is proposed in \cite{xu2016enhancing} to enhance secured coverage. The proposed scheme is based on the received signal power for the legitimate users where \acp{TP} far from the legitimate user are blocked for security and energy consumption concerns. The \ac{CoMP} scheme is also being used in \ac{UAV} systems to achieve secrecy.  \ac{CoMP} reception-enabled secrecy \ac{UAV} communication system is proposed in \cite{yao20193d}, in which multiple ground nodes cooperatively detect the legitimate information sent from the \ac{UAV} to enhance the legitimate communication performance in eavesdropper’s presence. 

Channel shortening, on the other hand, has been utilized to improve security against eavesdropping in \cite{furqan2017enhancing}, where the channel shortening equalizer is designed considering the channel between legitimate transmitter (Alice) and legitimate receiver (Bob). The security is provided by shortening the effective \ac{CIR} at Bob and designing/selecting the \ac{CP} accordingly. Since the \ac{CIR} experienced by the illegitimate eavesdropper (Eve) is longer, the signal it receives is prone to \ac{ISI}, deteriorating its reception. However, the shortening approach fails when Eve experiences a better channel (shorter \ac{CIR}) as compared to Bob. This shortcoming is addressed in the current work, which contributes the following:

\begin{itemize}
    \item This work proposes a \ac{CoMP}-assisted solution against eavesdropping which, in conjunction with channel shortening, addresses the limitation of multiple \ac{PLS} mechanisms that fail when the attacker experiences a better channel compared to the legitimate receiver.
    \item The effect of correlated channel between the eavesdropper and legitimate receiver on the proposed technique is also studied.
    \item The limitations and future directions for this work are provided.
\end{itemize}
The rest of this article is organized as follows. Section \ref{Sec:SystemModel} describes the preliminary system model and its associated assumption. The proposed method is given in Section \ref{Sec:ProposedApproach}. Section \ref{Sec:Simulation} provides the simulation results. Section \ref{Sec:Conclusion} concludes the paper by highlighting the limitations and future directions associated with this work.

\section{System Model and Assumptions}
\label{Sec:SystemModel}
In this work, downlink communication using \ac{OFDM} is considered. There are $K$ coordinating \acp{TP} present, that use \ac{JT} to serve the user in an urban environment. For simplicity, a single legitimate receiver, referred to as Bob, and a single eavesdropper called Eve are considered. Both Bob and Eve are located randomly within the coverage area of the cooperating \acp{TP} and are assumed to experience independent and uncorrelated frequency-selective Rayleigh fading (Section \ref{Sec:Simulation}, however, looks at the effect of correlated channels), albeit with the same exponentially decaying power delay profile. In general, the received signal at either Bob or Eve from the multiple cooperating \acp{TP} can be expressed as
\begin{equation}
\label{eq:rxSignal}
    Y_j(i)=\sum_{k=1}^{K}\sqrt{P_k}H_{j,k}(i)X_k(i)+ W(i),
\end{equation}
where $j \in \{b,e\}$, depending on whether the receiver is Bob or Eve, $X_k(i)$ is the $i$-th transmitted symbol from $k$-th \ac{TP}, $H_{j,k}(i)$ represents the $N$-point \ac{FFT} of the $k$-th \ac{TP}'s \ac{CIR}, $W(i) \sim \mathcal{N}(0,\,N_0B_T)$ represents the \ac{AWGN}, where $N_0$ is the noise \ac{PSD}, $B_T$ is the system bandwidth and $P_k$ is the signal power received from $k$-th \ac{TP}. This power is calculated considering the pathloss described for urban macrocell environment given in \cite{3GPP_38_901} using the following equation:
\begin{equation}
\label{pathloss}
    P_k=P_k^{Tx}-(22\log_{10}(d)+20\log_{10}(f_c)+28+\sigma),
\end{equation}
where $P_k^{Tx}$ is the power transmitted from the $k$-th \ac{TP}, $d$ is the 3-D distance between the \ac{TP} and the \ac{UE} in meters, $f_c$ is the carrier frequency in GHz, and $\sigma$ represents the shadow fading modeled as a zero-mean log-normal distribution. Since the goal of this work is to highlight the potential of the proposed method (described in Section \ref{Sec:ProposedApproach}) for securing the communication, we assume perfect channel estimation and ideal synchronization/backhaul \cite{3GPP_36_819} in the \ac{JT}-\ac{CoMP}. 

\begin{figure}[t!]
    \centering
    \includegraphics[width=\linewidth]{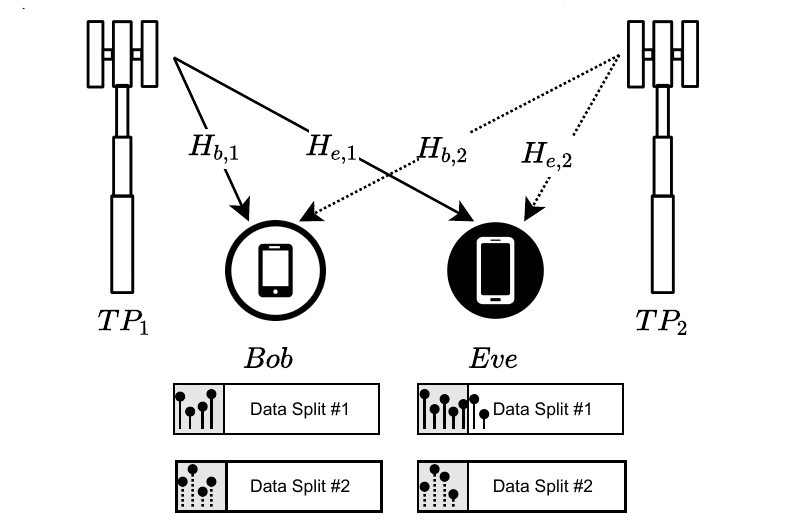}
    \caption{Illustration of the utilized system model for $K=2$ \acp{TP}. The \acp{TP} jointly transmit the data user data, which is received at both, Bob and Eve. $H_{j.k}$'s represent the channel over different links. The lower part of the figure shows the difference in delay spreads as experienced by both receivers (details in Section \ref{Sec:ProposedApproach}).}
    \label{fig:illustration_system}
\end{figure}

\begin{figure*}[t!]
    \centering
    \includegraphics[width=0.9\linewidth]{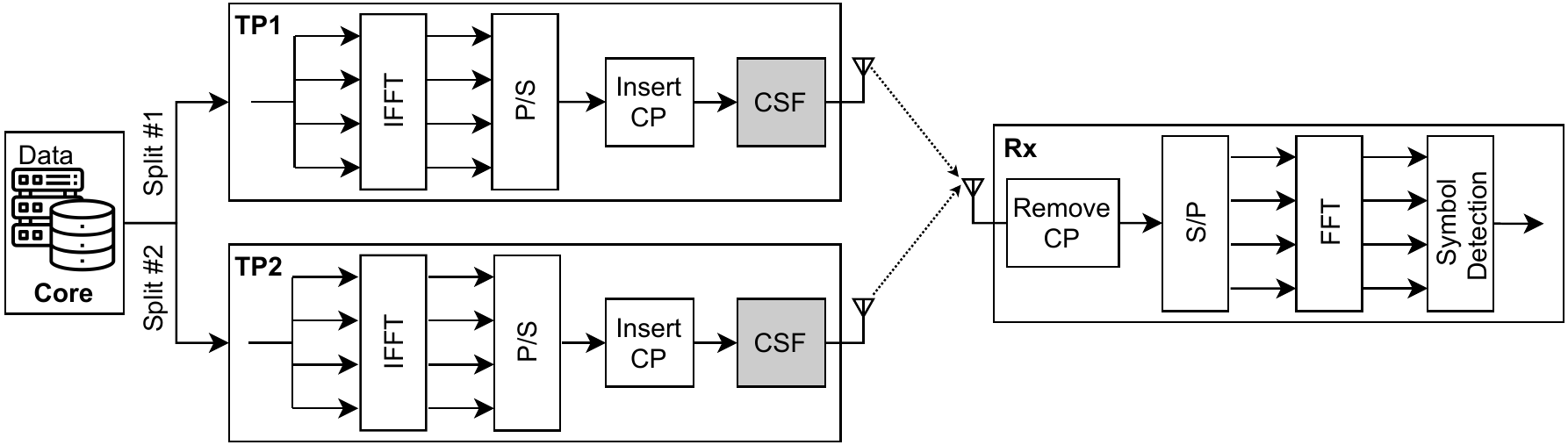}
    \caption{Illustration of the Tx/Rx blocks for the proposed method. The user data is split in to $K=2$ parts, where each part is sent to a distinct \ac{TP}. These \acp{TP} transmit their respective splits after employing \ac{CSF} which shortens the effective \ac{CIR}. At the Rx side, a typical \ac{OFDM} receiver is employed.}
    \label{fig:BlockDiagram}
\end{figure*}

\section{Proposed Approach}
\label{Sec:ProposedApproach}
\begin{figure}[t!]
    \centering
    \includegraphics[width=0.95\linewidth]{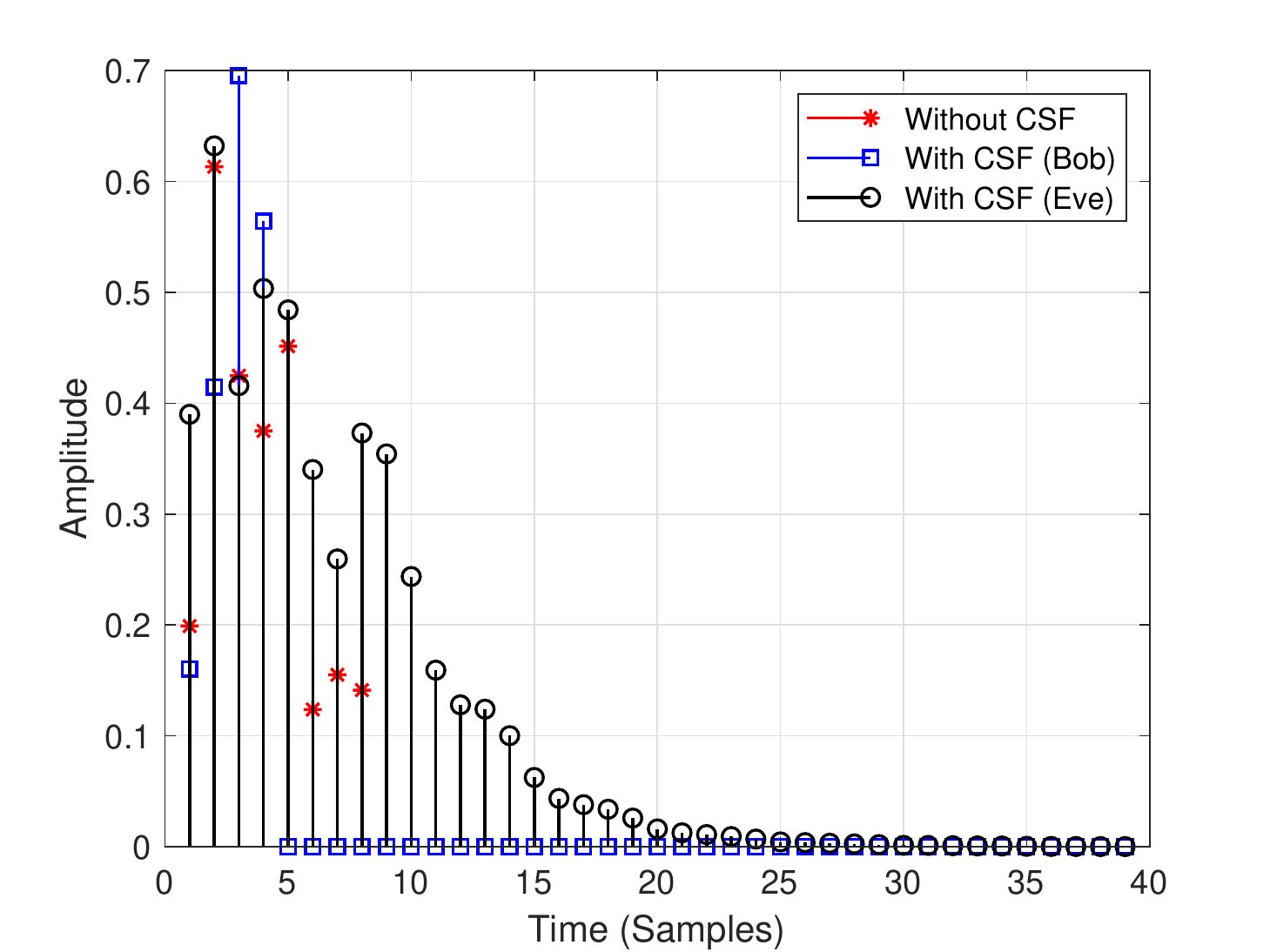}
    \caption{Effective CIR after (MSSNR) channel shortening is applied. Original CIR has a length of 8 taps, while the desired CIR length is 4 taps.}
    \label{fig:CIR_CSF}
\end{figure}

In this work, we aim to tackle the issue arising in conventional channel-adaptation based \ac{PLS} mechanisms when the channel of the eavesdropper is better as compared to the legitimate node. This is achieved by leveraging the geographically distributed \acp{TP} offered by \ac{CoMP} networks to provide security against eavesdropping attackers in a wireless communication system. The main stages of the proposed technique are listed below:
\begin{itemize}
    \item \textit{Selection of the coordinating \acp{TP}}:  Different criteria may be considered for the selection of coordinating \acp{TP}, such as the \ac{RSSI}, received power relative to the serving \ac{TP}, or a combination of both \cite{bassoy2016load, bassoy2019load}. It is also possible to have different sets of cooperating \acp{TP} for each user, or the same set for a group of users. The selection of coordination set can also be done for different objectives/constraints such as capacity maximization, spectral efficiency, energy efficiency, etc \cite{bassoy2017coordinated}. For the sake of this work, we have considered an interference-free scenario, where only two \acp{TP} are present and coordinating. Hence, there is no need of \textit{selecting} any \acp{TP}.
    \item \textit{Splitting of the data:} The data that is to be sent to the aforementioned user in split into $K$ parts, where $K$ is the number of \acp{TP} used. For the sake of this work, we restrict ourselves to a very simple approach, i.e., into real and imaginary parts where \ac{TP} 1 transmits the former, while \ac{TP} 2 sends the latter. Here it should be pointed out that further analysis is required to determine the optimum splitting mechanism, especially in the case of $K \neq 2$. However, it is beyond the scope of the present work.
    \item \textit{Channel-based manipulation:} This step covers the design of a channel-based manipulation mechanism for the transmitted signal to ensure the signal received at the legitimate node has delay spread, $\tau_{max}$ less than the guard duration, i.e., $\tau_{max} \leq T_g$. In other words \ac{CIR} length, $L$ (assuming sample spaced taps) is less than \ac{CP} length. However, the goal is to ensure that the same is NOT true for the illegitimate node, i.e., Eve. In our particular case, we consider the usage of a \ac{CSF}. A \ac{CSF} is typically used at the receiver to shorten the \ac{CIR} observed by it, allowing it to use smaller \ac{CP}/guard, thereby improving the spectral efficiency of the system \cite{melsa1996impulse}. However, it has been used for \ac{PLS} in \cite{furqan2017enhancing} as described in Section \ref{Sec:Intro} but is only applicable in the case where Eve is farther from the \ac{TP} compared to Bob. Since we use distributed \acp{TP} in this work, it ensures that Eve experiences longer $\tau_{max}$ over at least one of the links which will lead to \ac{isi} and degrade its interception capability. 
\end{itemize}
Figure \ref{fig:BlockDiagram} illustrates the block diagram for the proposed method. It can be seen that the data splitting and \ac{CSF} blocks set the transmitter apart from a conventional \ac{OFDM} one. On the other hand, receiver side is typical of what we expect in an \ac{OFDM} transceiver. The received signal can be mathematically represented by expanding \eqref{eq:rxSignal} as
\begin{equation}
\label{eq:rxSignalExpanded}
    Y_j(i)= \sqrt{P_1}H_{j,1}^{eff}(i)X_1(i) + \sqrt{P_2}H_{j,2}^{eff}(i)X_2(i) + W(i),
\end{equation}
where $X_1(i)$ and $X_2(i)$ represent the real and imaginary parts of the transmitted symbol, $X(i)$, i.e.,
\begin{equation}
    X_1(i)= \mathfrak{Re} \{X(i)\},
\end{equation}
\begin{equation}
    X_2(i)= \mathfrak{Im} \{X(i)\},
\end{equation}
and 
\begin{equation}
\label{eq:CSF}
    H_{j,k}^{eff}(i) = H_{j,k}(i)H_{b,k}^{csf}(i).
\end{equation}
Here, $H_{b,k}^{csf}$ refers to the \ac{MSSNR} \ac{CSF} \cite{melsa1996impulse} designed specifically for the link between the $k$-th \ac{TP} and Bob. It should be noted that the \ac{CSF} block is part of the transmitter, therefore, received signals at both Bob and Eve pass through the same shortening filter. However, since its design is done to increase the shortening \ac{SNR} for Bob only, the effective channel for both differ significantly, as shown in Fig. \ref{fig:CIR_CSF}. Original \ac{CIR} for both, Bob and Eve is assumed to have length, $L = 8$ while the desired \ac{CIR} length is set as $4$. The length of the \ac{MSSNR} shortening filter is four times the original \ac{CIR} length. It can be seen that the \ac{CIR} for Bob after shortening is indeed limited to $4$ taps (near zero values after that). On the other hand, for Eve a significant amount of energy lies outside the original \ac{CIR} length which is exploited in the form of \ac{ISI} to degrade its decoding capability. This is also shown in Fig. \ref{fig:illustration_system} where the \ac{CIR} experienced by Eve from the far \ac{TP} exceeds the \ac{CP} duration. For Bob, however, both links experience a \ac{CIR} which lies within the \ac{CP}.

\section{Simulation Results and Discussions}
\label{Sec:Simulation}
The simulations carried out in this works assume a macro urban environment, with two \acp{TP} at a distance of $500$m. Carrier frequency, $f_c$ is assumed to be $2$ GHz, and shadow fading standard deviation, $\sigma$ is $4$ dB. Total transmit power of the each \ac{TP} is $46$dBm and noise \ac{PSD}, $N_0$, is $-174$dBm/Hz. The original \ac{CIR} length, $L$ for both Bob and Eve is $8$. However, desired \ac{CIR} for the \ac{CSF} is $4$, which is also used as the \ac{CP} duration. The reported results are obtained over $2500$ iterations, where each iteration sends $64$ \ac{OFDM} symbols with \ac{QPSK} modulation. The receiver employs maximum likelihood detection to detect the real and imaginary parts of the symbols accurately. These and other simulation parameters are summarized in Table \ref{tab:parametersSec}.

\begin{figure}[t!]
    \centering
    \includegraphics[width=0.95\linewidth]{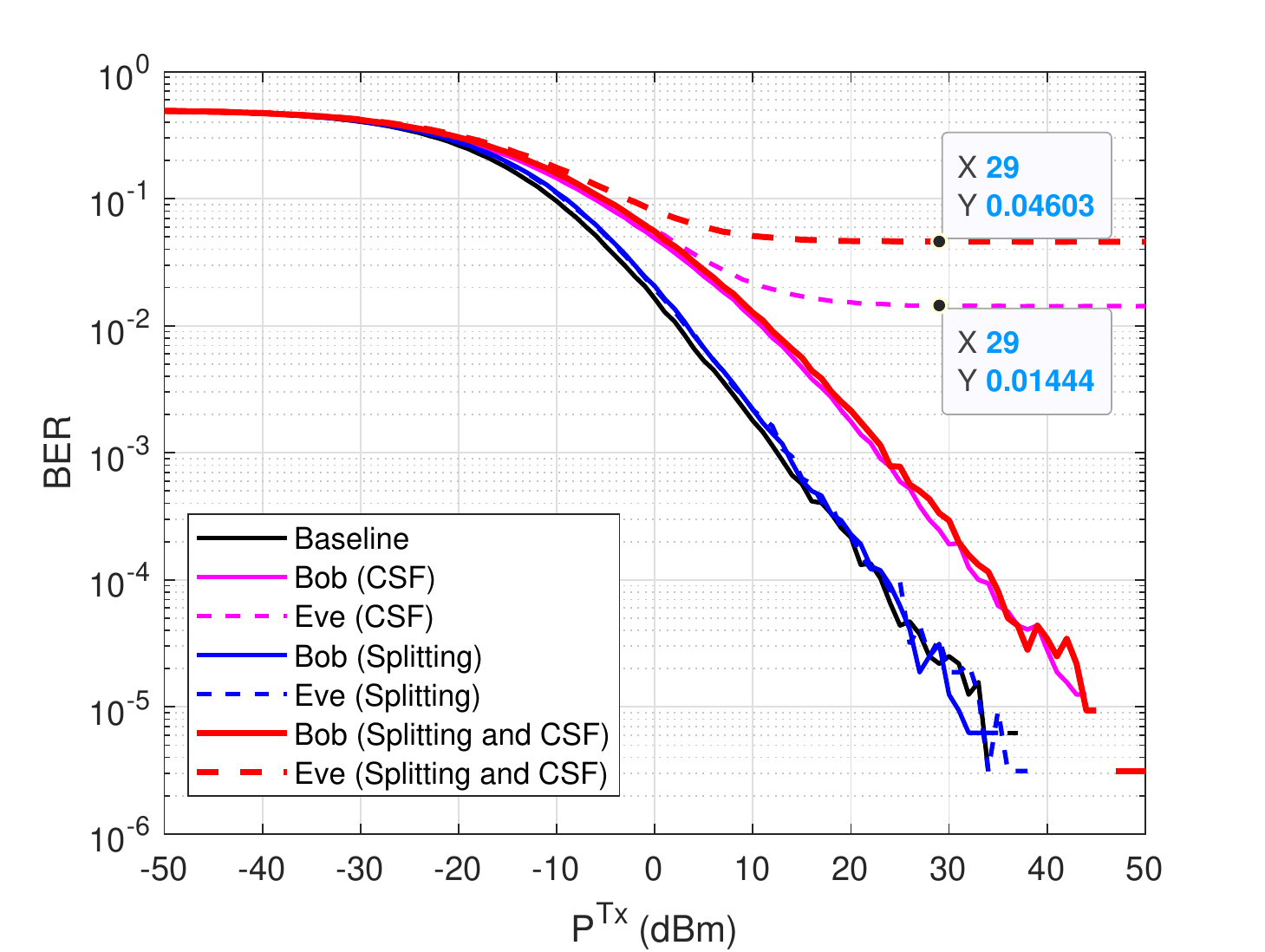}
    \caption{Performance comparison of the proposed approach with channel shortening, data splitting, and baseline OFDM in terms of BER at Bob and Eve.}
    \label{fig:BER_CoMP_Shorten}
\end{figure}

As mentioned earlier, the goal of Eve is to intercept and interpret the ongoing legitimate communication. Consequently, the target of any \ac{PLS} mechanism is to ensure that Eve is unable to capture the least possible information. A practical metric used to evaluate the performance of the security mechanisms, therefore, is the gap in error rates (bit, symbol, packet, etc.) experienced by Bob and Eve, which is also referred to as the security gap. Ideally, this should be increased without compromising Bob's (baseline) performance. Accordingly, in this work we evaluate the performance of the proposed approach by comparing the \acp{BER} experienced by both Bob and Eve when i) only channel shortening is used, ii) only data splitting is used, iii) both shortening and splitting are used, as shown in Fig. \ref{fig:BER_CoMP_Shorten}. The \ac{BER} performance of a conventional \ac{OFDM} system is also presented to serve as a baseline. 

\hfill

\begin{figure}[t!]
    \centering
    \includegraphics[width=0.95\linewidth]{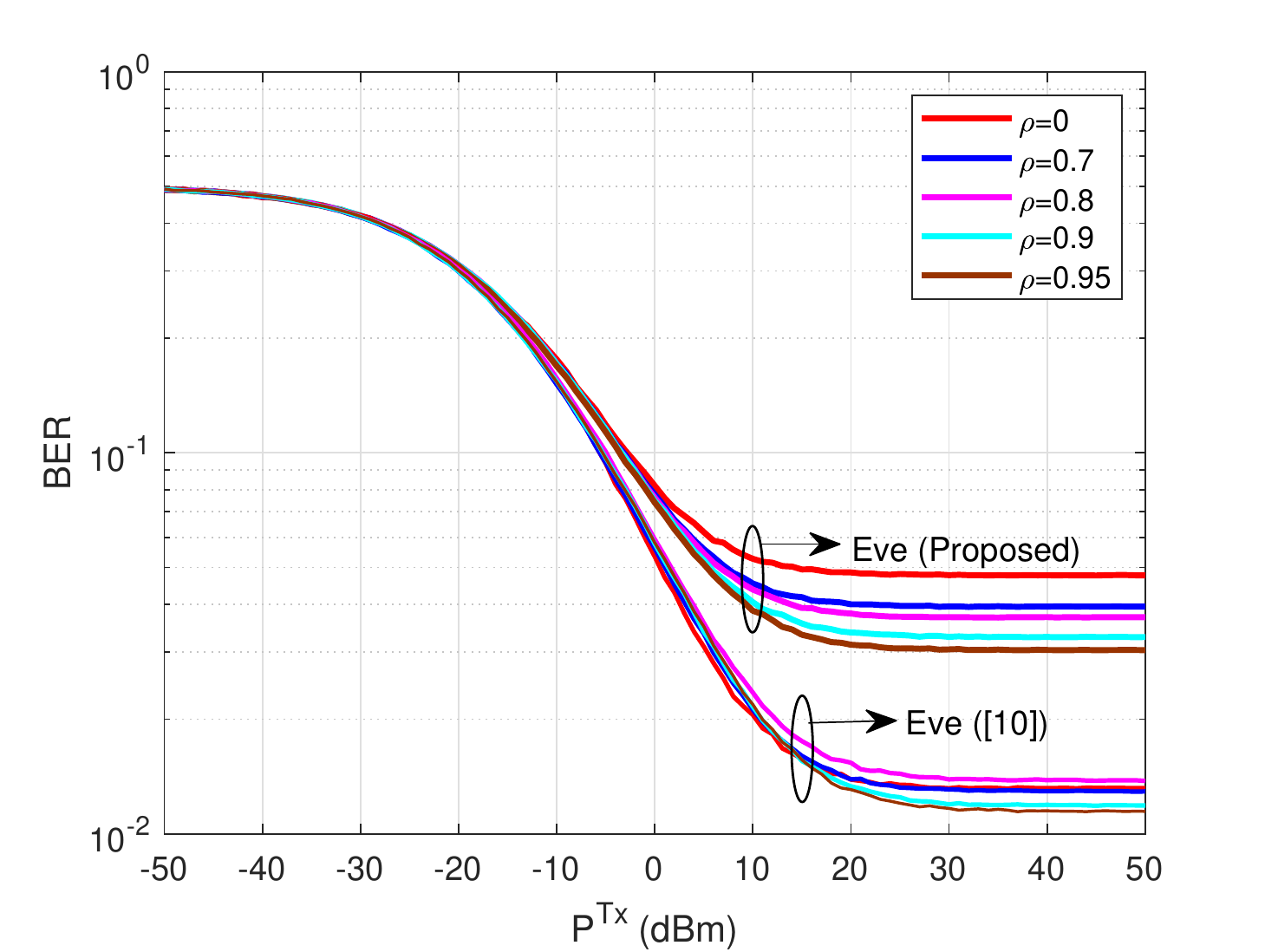}
    \caption{Performance comparison of the proposed approach with channel shortening \cite{furqan2017enhancing} in the presence of channel correlation between Bob and Eve, where $\rho$ represents the correlation coefficient.}
    \label{fig:ChannCorr}
\end{figure}

\begin{table}[t!]
\centering
\caption{Simulation parameters and assumptions} \label{tab:parametersSec}
\resizebox{0.80\columnwidth}{!}{%
\begin{tabular}{|l|l|}
\hline
\textbf{Parameter}      & \textbf{Value}                \\ \hline
Simulation environment             & Urban macro                   \\ \hline
Carrier frequency $(f_c)$     & 2 GHz                         \\ \hline
Number of RBs/TP       & 50               \\ \hline
System bandwidth $(B_T)$        & 10 MHz               \\ \hline
Shadow fading standard deviation $(\sigma)$        & 4 dB               \\ \hline
Noise power density $(N_0)$     & -174 dBm/Hz                    \\ \hline
Inter-site distance     & 500 m 
\\ \hline
CIR length ($L$)      & 8 
\\ \hline
Desired CIR ($= N_{CP}$ )      & 4 
\\ \hline
\end{tabular}%
}
\end{table}

\begin{table*}[htp!]
\centering
\caption{BER comparison of the proposed method for different symbol and CP durations}
\label{tab:BER_CoMP_Shorten}
\resizebox{0.7\textwidth}{!}{%
\begin{tabular}{|c|c|c|c|c|c|}
\hline
\multirow{2}{*}{$\textbf{N}_{\textbf{Data}}$} & 
\multirow{2}{*}{\begin{tabular}[c]{@{}c@{}}$\textbf{N}_{\textbf{CP}}$\\ \textbf{(Desired CIR)}\end{tabular}}
& \multirow{2}{*}{\textbf{BER - Bob \cite{furqan2017enhancing}}} & \multirow{2}{*}{\begin{tabular}[c]{@{}l@{}}\textbf{BER - Bob}\\ \textbf{(Proposed)}\end{tabular}} & \multirow{2}{*}{\textbf{BER - Eve \cite{furqan2017enhancing}}} & \multirow{2}{*}{\begin{tabular}[c]{@{}l@{}}\textbf{BER - Eve} \\ \textbf{(Proposed)}\end{tabular}} \\
 & & &                                                                                     & &  \\ \hline
\multirow{5}{*}{64}   & 4    & 0.00025   & 0.00033  & 0.01444   & 0.04603   \\ \cline{2-6} 
                      & 5   & 0.00061   & 0.00065  & 0.01046   & 0.03941  \\ \cline{2-6} 
                      & 6   & 0.00079   & 0.00085  & 0.00843   & 0.02924   \\ \cline{2-6} 
                      & 7   & 0.00108   & 0.00118  & 0.00712   & 0.02544   \\ \cline{2-6} 
                      & 8   & 0.00127   & 0.00138  & 0.00609   & 0.02270    \\ \hline
\multirow{2}{*}{128}  & 8   & 0.00146   & 0.00159  & 0.00324   & 0.01501   \\ \cline{2-6} 
                      & 16   & 0.00047   & 0.00049  & 0.03026   & 0.09628   \\ \hline
\end{tabular}}
\end{table*}

It can be seen that data splitting itself provides no advantage in terms of security as both Bob and Eve experience similar \acp{BER} as the baseline. This is expected since they are like the baseline case themselves when looked at in isolation. As in the case of \cite{furqan2017enhancing}, channel shortening provides a significant gain in terms of the security gap while providing a degraded \ac{BER} performance for Bob as compared to the baseline case. The proposed approach improves on the shortening, by further exacerbating Eve's performance without any degradation in Bob's reception quality. Quantitatively speaking, there is approximately threefold worsening of Eve's \ac{BER} at $29$ dBm transmit power, as shown in Fig. \ref{fig:BER_CoMP_Shorten}. This specific value is chosen since it represents the transmit power for one \ac{RB} when the maximum transmit power is equally distributed amongst all \acp{RB}. Since \ac{CSF} performance is closely related to the original and desired \ac{CIR} lengths, we have evaluated the system for various symbol ($N_{Data}$) and \ac{CP} ($N_{CP}$) durations where the \ac{CP} length is always chosen as equal to the desired \ac{CIR} length, which is in turn, half the original \ac{CIR}. Table \ref{tab:BER_CoMP_Shorten} summarizes these results. It can be seen that as the \ac{CP} (and desired \ac{CIR}) ratio increases, the gap in \ac{BER} performance of Bob and Eve reduces significantly. This indicates the importance of selecting the most appropriate \ac{CP}/desired \ac{CIR} length during the shortening process. 

Here, it should be noted that we have assumed the channels between the \acp{TP} and Bob and Eve to be uncorrelated and independent. In fact, most channel-based \ac{PLS} mechanisms rely on this assumption to provide the required security gap. Accordingly, we look at the performance of the proposed method in the presence of correlation between Eve and Bob's channels. The correlated channel for Eve can be mathematically represented as \cite{ferdinand2013physical}
\begin{equation}
\label{eq:channCorr}
    H_{e,k}^{corr}= \rho H_{b,k} + \sqrt{1 - \rho^2}H_{e,k},
\end{equation}
where $\rho$ is correlation coefficient. Figure \ref{fig:ChannCorr} compares Eve's performance for different values of $\rho$ for the proposed method, and also with the technique provided in \cite{furqan2017enhancing}. While channel correlation has a degrading effect on both methods, it is interesting to note the even at a high correlation ($\rho = 0.95$), the proposed method gives significantly worse \ac{BER} for Eve (and better security) as compared to the to the best case ($\rho = 0$) for the channel shortening based method using a single \ac{TP} \cite{furqan2017enhancing}.

\section{Conclusion and Future Work}
\label{Sec:Conclusion}
Enhancing the privacy and security of user data is increasingly becoming a major concern for wireless networks. Conventional methods such as cryptography are prone to computationally advanced attackers, while also suffering from key management issues, especially in ultra-dense heterogeneous networks. \ac{PLS} provides a complementary solution to these security concerns, however, the majority of these methods also struggle when the eavesdropper boasts a better (stronger) channel as compared to the legitimate receiver. In this work, the distributed \acp{TP} offered by \ac{CoMP} are leveraged to address this problem. Specifically, the user data is split into multiple parts, where each part is sent using a different \ac{TP}. This ensures that at least one of the illegitimate links has a worse channel as compared to the legitimate one. Additionally, a channel shortening filter is applied w.r.t legitimate links, which results in \ac{ISI} being introduced at the receiver. Simulation results verify that the proposed method is more advantageous as compared to the existing shortening-based \ac{PLS} mechanism. 

It should be noted that this work provides a rudimentary method (and analysis) of leveraging \ac{CoMP} with channel shortening for security. Further studies are required to fully explore the potential of these two mechanisms in securing communication. In this context, one of the future studies that we believe is highly merited is the analysis of the effect of data splitting mechanisms and its dependence on the number of cooperating \acp{TP} in improving security. Furthermore, it might also be worthwhile to look at the performance of the proposed method (and any other mechanisms arriving from it) under different propagation environments, \ac{LoS} assumptions, channel estimation errors, and scattering levels, etc.

\section*{Acknowledgment}
The authors would like to thank Haji M. Furqan for his valuable input during the preparation of this manuscript.

\bibliographystyle{IEEEtran}


\end{document}